# Nanoscale Metamaterial Optical Waveguides with Ultrahigh Refractive Indices

Yingran He[1,2], Sailing He[2], Jie Gao[1,*] and Xiaodong Yang[1,†]

[1] *Department of Mechanical and Aerospace Engineering, Missouri University of Science and Technology, Rolla, MO 65409, USA*

[2] *Centre for Optical and Electromagnetic Research, Zhejiang Provincial Key Laboratory for Sensing Technologies, Zhejiang University, Hangzhou 310058, China*

*Corresponding author: [*]gaojie@mst.edu, [†]yangxia@mst.edu*

**Abstract:** We propose deep-subwavelength optical waveguides based on metal-dielectric multilayer indefinite metamaterials with ultrahigh effective refractive indices. Waveguide modes with different mode orders are systematically analyzed with numerical simulations based on both metal-dielectric multilayer structures and the effective medium approach. The dependences of waveguide mode indices, propagation lengths and mode areas on different mode orders, free space wavelengths and sizes of waveguide cross sections are studied. Furthermore, waveguide modes are also illustrated with iso-frequency contours in the wave vector space in order to investigate the mechanism of waveguide mode cutoff for high order modes. The deep-subwavelength optical waveguide with a size smaller than $\lambda_0/50$ and a mode area in the order of $10^{-4} \lambda_0^2$ is realized, and an ultrahigh effective refractive index up to 62.0 is achieved at the telecommunication wavelength. This new type of metamaterial optical waveguide opens up opportunities for various applications in enhanced light-matter interactions.

**Keywords:** Subwavelength optical waveguide, optical metamaterial, plasmonics, ultrahigh refractive index

The emergence of metamaterials with artificially engineered subwavelength composites offers a new perspective on light manipulation and exhibits intriguing optical phenomena such as negative refraction [1-3], sub-diffraction imaging [4-6], invisible cloaking [7-9] and high index of refraction [10-12]. One unique kind of optical metamaterials is the indefinite metamaterial with extreme anisotropy, in which not all the principal components of the permittivity tensor have the same sign [13]. The non-magnetic design and the off-resonance operation of the indefinite metamaterial can considerably reduce the optical absorption associated with conventional metamaterials [14]. The unique hyperbolic dispersion of the indefinite metamaterial enables the demonstration of negative refraction [15], sub-diffraction optical imaging with hyperlenses [4-6], the strong enhancement of photonic density of states [16, 17], slow-light waveguides [18-20] and broadband light absorbers [21]. Since the hyperbolic dispersion eliminates the cutoff of large wave vectors, the effective refractive index can be arbitrarily high. Such a capability is potentially important for building nanoscale optical cavities [22] and deep-subwavelength optical waveguides [23, 24] with strong optical energy confinement. In reality, besides natural indefinite media such as graphite in the ultraviolet spectrum [25], an indefinite metamaterial is usually constructed with metal-dielectric multilayer structures [6] rather than metallic nanowires array embedded in dielectrics [15] due to the difficulties of device fabrication.

Since optical waveguides play an important role in many fundamental studies of optical physics at nanoscale and in exciting applications in nanophotonics, optical waveguides based on indefinite metamaterials have been recently studied [19, 20, 23, 24, 26-29], in order to obtain novel optical properties beyond the conventional dielectric waveguides, especially slow light propagation [19, 20], surface modes guidance [28, 29], as well as subwavelength mode compression [23, 24]. In this paper, we propose deep-subwavelength optical waveguides based on metal-dielectric multilayer indefinite metamaterials, which support waveguide modes with tight photon confinement due to the ultralarge wave vectors inside indefinite metamaterials, and therefore ultrahigh effective refractive indices. The optical properties of waveguide modes will be presented with the mode analysis in both the real space and the wave vector space. The dependences of waveguide mode indices, propagation lengths and mode areas on different mode orders, free space wavelengths and sizes of waveguide cross sections are investigated. The mechanism of waveguide mode cutoff is also illustrated in the wave vector space. This new type of metamaterial optical waveguide with ultrahigh refractive indices and extremely tight photon confinement will be of great importance in the enhancement of light-matter interactions, such as nanoscale lasers [30], quantum electrodynamics [31], nonlinear optics [32], optomechanics [33], and transformation optics [34].

Fig. 1(a) shows the schematic of optical waveguides based on metal-dielectric multilayer indefinite metamaterials. The multilayer metamaterial is constructed with alternative layers of silver (Ag) and germanium (Ge). Each period includes a 4 nm silver layer and a 6 nm germanium layer, leading to a multilayer structure with period $a$ = 10 nm and filling ratio of silver $f_m$ = 0.4. If the period of the multilayer $a$ is much less than the operation wavelength ($a << 2\pi/k$, where $k$ is wave vector), the multilayer metamaterial can be treated as a homogeneous effective medium and the principle components of the anisotropic permittivity tensor can be determined from the Maxwell-Garnet theory [35, 36],

$$\varepsilon_x = \varepsilon_z = f_m \varepsilon_m + (1 - f_m) \varepsilon_d$$
$$\varepsilon_y = \frac{\varepsilon_m \varepsilon_d}{f_m \varepsilon_d + (1 - f_m) \varepsilon_m} \quad (1)$$

where $f_m$ is the volume filling ratio of silver, $\varepsilon_d$ and $\varepsilon_m$ are the permittivity corresponding to germanium and silver, respectively. The permittivity of germanium is $\varepsilon_d$ = 16, and the optical properties of silver are described by the Drude model $\varepsilon_m(\omega) = \varepsilon_\infty - \omega_p^2/(\omega^2 - i\omega\gamma)$, with a background dielectric constant $\varepsilon_\infty$ = 5, plasma frequency $\omega_p = 1.38 \times 10^{16}$ rad/s and collision frequency $\gamma = 5.07 \times 10^{13}$ rad/s [37]. Fig.

1(b) shows the calculated effective permittivity tensor for the multilayer metamaterial with a silver filling ratio of $f_m = 0.4$ for the free space wavelength $\lambda_0$ ranging from 1 μm to 2 μm. The permittivity shows negative value along $x$ and $z$ directions (parallel to the multilayers) and positive value along $y$ direction (vertical to the multilayers). Since large wave vectors are supported in indefinite metamaterials due to the hyperbolic dispersion, ultrahigh refractive indices can be reached, which will enable the formation of optical waveguides with deep subwavelength cross sections based on the total internal reflection (TIR) at the interface between metamaterial and air, as illustrated in Fig. 1(a).

Fig. 2 plots the waveguide mode profiles of different mode orders supported in a metamaterial waveguide with a cross section of $L_x = L_y = 100$ nm at free space wavelength $\lambda_0 = 1$ μm, calculated from the finite-element method (FEM) software package (COMSOL). The distributions of optical field components $H_x$, $E_y$, and electromagnetic (EM) energy density $W$ for different modes calculated from the multilayer structure are shown in Fig. 2 (a), (c) and (d). It is clear that these waveguide modes with different mode orders $(m_x, m_y)$ exhibit spatial oscillations in both $x$ and $y$ directions, with specified wave vectors $k_x$ and $k_y$, in addition to the propagation wave vector $k_z$ along the waveguide. For the mode order $(m_x, m_y)$, $m_x$ and $m_y$ represent the number of peaks in $H_x$ profile inside the waveguide along the $x$ direction and the $y$ direction, respectively. Compared with the $H_x$ components, the $E_y$ components show exactly the same number of peaks within the waveguide. However, the $W$ profile is related to the mode order $(m_x, m_y)$ in a different way, and its maximums approximately occur at locations where $H_x$ and $E_y$ have the largest spatial gradients. The EM energy density $W$ is calculated by taking the strongly dispersive property of silver into account, as $W(x,y) = \frac{1}{2}\text{Re}\left[\frac{d(\omega\varepsilon_m)}{d\omega}\right]\varepsilon_0|\vec{E}|^2 + \frac{1}{2}\mu_0|\vec{H}|^2$. Fig. 2(b) gives the distributions of optical field components for $(1, m_y)$ mode calculated with the effective medium method, which agree very well with the multilayer results in Fig. 2(a).

In order to study the optical properties of these waveguide modes with different mode orders in the waveguide with a cross section of $L_x = L_y = 100$ nm, waveguide mode indices along the propagation direction $n_{\text{eff},z}$, propagation lengths $L_m$ and mode areas $A_m$ as functions of free space wavelength $\lambda_0$ are shown in Fig. 3. As indicated from Fig. 3(a), waveguide modes with a lower $m_x$ or a higher $m_y$ tend to have larger mode indices $n_{\text{eff},z}$ ($= k_z/k_0$) at a specific wavelength $\lambda_0$. As a result, the mode index of the (1, 1) mode is the largest in the ($m_x$, 1) mode group and the smallest in the (1, $m_y$) mode group. It is clear that waveguide mode indices will decrease as the wavelength $\lambda_0$ grows, which is caused by the material dispersion shown in Fig. 1(b), but the decrease rates depend on different mode orders $m_x$. Waveguide mode with a higher mode order $m_x$ along the $x$ direction will have a faster index decrease rate due to the larger $k_x$ and the stronger effect from the material dispersion, which will eventually result in the waveguide mode cutoff ($n_{\text{eff},z} < 1$) at a lower wavelength. In contrast to the (3, $m_y$) and (2, $m_y$) modes, the (1, $m_y$) modes do not have mode cutoff due to the zero wave vector along the $x$ direction ($k_x = 0$). The propagation lengths calculated from $L_m \equiv 1/2\,\text{Im}(k_z) = \lambda_0 / 4\pi\,\text{Im}(n_{\text{eff},z})$ are plotted in Fig. 3(b). When the wavelength is much lower than the mode cutoff wavelength, higher order modes for a given wavelength have shorter propagation lengths, since tight mode confinement in high order modes will induce large absorption loss. As an example, for the (2, 1) mode, as the wavelength increases, the propagation length will go up first, due to the reduced absorption loss in the material dispersion. When the wavelength gets close to the mode cutoff wavelength, the propagation length will drop dramatically, due to the increased radiation leakage of the waveguide mode confined by TIR. The observed behavior of the propagation length results from the tradeoff between the absorption loss and the mode radiation loss. The (1, $m_y$) modes maintain long propagation lengths since there is no mode cutoff. The (1, 1) mode turns out to have the longest propagation length, which is around 700 nm at $\lambda_0 = 1.55$ μm. This subwavelength propagation length in fact covers several operation wavelengths inside the waveguide with a high mode index of 8.3 at $\lambda_0 = 1.55$ μm. Fig. 3(c) gives the calculated mode areas $A_m$ for all the waveguide modes, where $A_m \equiv \dfrac{\iint W(x,y)\mathrm{d}x\mathrm{d}y}{\max[W(x,y)]}$. The results show that the metamaterial waveguides give deep-subwavelength optical energy confinement with the mode areas down to the order of $10^{-3}$ ($\lambda_0^2$).

Since the multilayer metamaterial can be treated as an indefinite medium with the principle components of permittivity tensor calculated in Eq. (1), the dispersion relation for such uniaxial anisotropic material is [22]:

$$\frac{k_x^2 + k_z^2}{\varepsilon_y} + \frac{k_y^2}{\varepsilon_z} = k_0^2 \tag{2}$$

where $k_0$ is the free space wave vector corresponding to the free space wavelength $\lambda_0$. The components of effective refractive index $n_{\text{eff}}$ are related to wave vector components along different directions, as $(n_{\text{eff},x}, n_{\text{eff},y}, n_{\text{eff},z}) = (k_x/k_0, k_y/k_0, k_z/k_0)$. The 3D hyperboloid iso-frequency contour (IFC) of indefinite metamaterial in $k$-space for a specific $\lambda_0$ is shown in Fig. 4(a), where the wave vector component along the propagation direction $k_z$ can be obtained from Eq. (2) if the other two components $k_x$ and $k_y$ are known for a specific mode order $(m_x, m_y)$. As shown in Fig. 2, the $H_x$ field distributions show harmonic oscillations with a cosine function along the $x$ direction and a sine function along the $y$ direction, due to the extreme anisotropy of the indefinite metamaterial and therefore different boundary conditions on the waveguide interfaces. For the (1, 1) mode, the field profile shows a zero phase accumulation (a constant phase) along the $x$ direction, and a $\pi$ phase accumulation along the $y$ direction. In general, for a specific $(m_x, m_y)$ mode, $H_x$ will undergo a phase accumulation of $(m_x-1)\pi$ along the $x$ direction and a phase accumulation of $m_y\pi$ along the $y$ direction. The values of $k_x$ and $k_y$ are then related to the size of the waveguide cross section and the mode order,

$$\begin{aligned} k_x &= (m_x - 1)\frac{\pi}{L_x} \\ k_y &= m_y \frac{\pi}{L_y} \end{aligned} \tag{3}$$

with $m_x$ = 1, 2, 3 and $m_y$ = 1, 2, 3 for the waveguide modes shown in Fig. 2. Based on the values of the wave vector components, a waveguide mode can be mapped on the hyperboloid surface in the 3D $k$-space. If there are common crossing points between two perpendicular cutting planes as defined in Eq. (3) and the 3D hyperboloid surface, the waveguide mode with a mode order of ($m_x$, $m_y$) will exist. Next, waveguide modes with different mode orders will be illustrated with IFCs in the $k$-space in order to investigate the mechanism of waveguide mode cutoff for high order modes.

In Fig. 4, waveguide modes supported in the waveguide with fixed size of $L_x$ = 100 nm and $L_y$ = 100 nm are analyzed in $k$-space. Fig. 4(a) shows the hyperboloid IFC for a specific wavelength $\lambda_0$, together with two cutting planes in grey color at $k_x$ = 0 and $k_x/k_0 = \lambda_0/L_x$ which represent mode orders of $m_x$ = 1 and $m_x$ = 3, respectively. The crossing curves between the two cutting planes and the hyperboloid IFC surface are plotted in Fig. 4(b) and Fig. 4(c), respectively, at different wavelengths of $\lambda_0$ = 1 μm (blue color) and $\lambda_0$ = 1.55 μm (red color). These hyperbolic curves are calculated from the effective medium method. The shapes of the curves will change as the wavelength varies, due to the permittivity dispersion of the indefinite metamaterial at different frequencies. The markers in Fig. 4(b) and Fig. 4(c) represent the (1, $m_y$) and (3, $m_y$) modes calculated from multilayer metamaterial waveguide structures, which locate on the IFCs showing that the effective medium approximation is valid in this situation. It is noted that the hyperbolic curves have changed their opening directions from the $z$ direction in Fig. 4 (b) to the $y$ direction in Fig. 4(c). According to Fig. 4(b), there are no mode cutoff for the (1, $m_y$) modes. However, the (3, 1) and (3, 2) modes have cutoff at $\lambda_0$ = 1.55 μm in Fig. 4(c), since $k_z$ is not available for the given mode orders. For $\lambda_0$ = 1 μm, the dispersion curve has much lower $k_y$, so that all the (3, $m_y$) modes still exist.

The effects of waveguide cross sections on the mode cutoff are then studied at a fixed wavelength of $\lambda_0$ = 1.55 μm. By varying the waveguide height $L_y$ or the waveguide width $L_x$, wave vectors will change based on Eq. (3), so that the corresponding cutting planes will also shift. Fig. 5 illustrates the waveguide modes in $k$-space for two waveguides with the same width $L_x = 100$ nm but different heights, $L_y = 80$ nm and $L_y = 150$ nm, respectively. In Fig. 5(a), three cutting planes corresponding to $m_x = 1$, 2 and 3 for the fixed $L_x$ are drawn in red, green and blue, respectively. The crossing curves between these cutting planes and the hyperboloid IFC surface are plotted in Fig. 5(b) and Fig. 5(c) for different waveguide heights, which are calculated from the effective medium method. It is noted that the hyperbolic curves with $m_x = 1$ open towards the $z$ direction, while the hyperbolic curves with $m_x = 2$ and $m_x = 3$ open towards the $y$ direction. The markers in Fig. 5(b) and Fig. 5(c) represent the ($m_x$, $m_y$) modes calculated from multilayer metamaterial waveguide structures. All the (1, $m_y$) modes located on a hyperbolic curve with $z$-direction opening do not have mode cutoff since there is always a corresponding $k_z$, no matter what the value of $k_y$ is. However, as shown in Fig. 5(b), the (3, 1) mode has cutoff for $L_y = 80$ nm since $k_z$ is not available for the given mode order. In Fig. 5(c), all the (3, $m_y$) modes and the (2, 1) mode have cutoff due to the reduced $k_y$ value as $L_y$ gets larger. Fig. 6 plots the waveguide modes in $k$-space for two waveguides with the same height $L_y = 100$ nm but different widths, $L_x = 120$ nm and $L_x = 70$ nm, respectively. In Fig. 6(a), three cutting planes corresponding to $m_y = 1$, 2 and 3 for the fixed $L_y$ are drawn in red, green and blue, respectively. The crossing circles between the cutting planes and the hyperboloid IFC surface are plotted in Fig. 6(b) and Fig. 6(c) for different waveguide widths. The circular IFC in $k_x$-$k_z$ plane implies that a waveguide mode will exist if $k_x$ is less than the radius of the circle. In Fig. 6(b), the (3, 1) mode has cutoff when $L_x = 120$ nm. In Fig. 6(c), all the (3, $m_y$) modes and the (2, 1) mode have cutoff, due to the increased $k_x$ value as $L_x$ gets smaller. In addition, all the (1, $m_y$) modes are not sensitive to the change of the waveguide width, due to $k_x = 0$. According to the above analysis in Fig. 5 and Fig. 6, a metamaterial waveguide with a small width $L_x$ and a large height $L_y$ (equivalent to a large $k_x$ and a small $k_y$) tend to have mode cutoff for high order modes at a fixed wavelength.

As indicated by the above analysis, the deep-subwavelength waveguides made of indefinite metamaterial support ultralarge wave vector components in both the propagation direction ($k_z$) and the lateral direction ($k_x$ and $k_y$), due to the unbounded hyperbolic dispersion. It will result in ultrahigh effective refractive indices $n_{eff}$, which is defined as:

$$n_{eff} = \sqrt{n_{eff,x}^2 + n_{eff,y}^2 + n_{eff,z}^2} = \sqrt{\left(\frac{k_x}{k_0}\right)^2 + \left(\frac{k_y}{k_0}\right)^2 + \left(\frac{k_z}{k_0}\right)^2} \quad (4)$$

Fig. 7(a) presents the dependence of effective refractive indices $n_{eff}$ for the (1, $m_y$) modes on waveguide sizes $L$ for waveguides with square cross sections ($L_x = L_y = L$) at $\lambda_0 = 1.55$ μm. The effective refractive index will increase as the waveguide size shrinks and the mode order $m_y$ gets higher. For example, $n_{eff} = 56.6$ is obtained for the (1, 3) mode for a waveguide with $L = 50$ nm, and $n_{eff} = 62.0$ is achieved for the (1, 2) mode for a waveguide with $L = 30$ nm. For $L = 30$ nm, the (1, 3) mode is not available from the multilayer waveguide structure calculation, since the period $a$ (= 10 nm) of the multilayer metamaterial structure is larger than the operation wavelength. In fact, the waveguide mode in the metal-dielectric multilayer structure is intrinsically the evolution of coupled metal-dielectric-metal (MDM) plasmonic modes [38]. For the multilayer structure with a period of 10 nm, there is only three silver layers and thus two coupled MDM plasmonic modes, so that only two modes [the (1, 1) and (1, 2) modes] are supported. The wave vectors of the waveguide modes shown in Fig. 7(a) are plotted in $k$-space in Fig. 7(b), which match the effective medium calculation (solid curve) in the low $k$ region ($k \ll 2\pi/a$). In the high $k$ region (when $L \leq 50$ nm), as the wavelength is close to the period of the multilayer structure $a$, the wave vectors calculated from the multilayer structure deviate from the effective medium simulation at the edge of the first Brillouin zone. It is demonstrated that ultralarge wave vectors are supported with the multilayer metamaterial waveguide structures in both $y$ and $z$ directions, leading to ultrahigh effective refractive indices $n_{eff}$ in such materials.

Finally, mode propagation of the (1, 1) modes inside the metamaterial waveguides is studied in 3D real space. Fig. 8(a) and Fig. 8(b) plot the distributions of magnetic field $H_x$ for the (1, 1) modes at $\lambda_0 = 1.55$ μm in waveguides with square cross sections of 30 nm by 30 nm, and 50 nm by 50 nm, respectively. Although the length of waveguides along the *z* direction $L_z$ is only 300 nm, the guided waves can undergone several oscillations before they are completely absorbed, as a result of the ultralarge waveguide mode indices. The waveguide mode properties are listed at the bottom of each figure. The tradeoff between the mode confinement and the propagation length is shown clearly, where the waveguide with $L = 30$ nm has a smaller mode area but a shorter propagation length, compared to the waveguide with $L = 50$ nm.

In conclusion, we have demonstrated nanoscale optical waveguides made of metal-dielectric multilayer indefinite metamaterials supporting ultrahigh effective refractive indices. Numerical simulation based on both metal-dielectric multilayer structures and the effective medium approach is performed systematically to analyze waveguide mode properties for different mode orders, including waveguide mode indices, propagation lengths, mode areas, and effective refractive indices. The mechanism of waveguide mode cutoff for high order modes is revealed with iso-frequency contours in the wave vector space. The deep-subwavelength optical waveguide with size less than $\lambda_0/50$ and mode area in the order of $10^{-4} \lambda_0^2$ is realized, and an ultrahigh effective refractive index up to 62.0 is achieved at $\lambda_0 = 1.55$ μm. These ultra-compact metamaterial waveguides opens a new realm of the enhanced light-matter interactions for many promising applications.

**Acknowledgements**

This work was partially supported by the Department of Mechanical and Aerospace Engineering, the Materials Research Center (MRC), the Intelligent Systems Center (ISC) at Missouri S&T, the University of Missouri Research Board, and the National

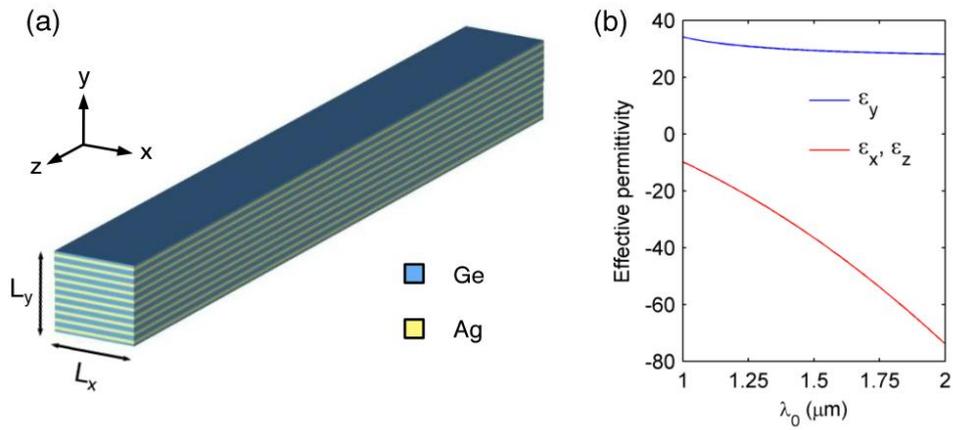

**Figure 1.** (a) Schematic of a nanoscale optical waveguide made of the metal-dielectric multilayer indefinite metamaterial, which is constructed with alternative layers of 4 nm silver and 6 nm germanium. Nanoscale optical waveguides can be created due to the TIR at the metamaterial-air interface. (b) The principal components of the permittivity tensor for the multilayer metamaterial with a silver filling ratio of $f_m = 0.4$, calculated from the effective medium theory (Eq. 1), where $\varepsilon_y$ is positive and $\varepsilon_z$ ($\varepsilon_x$) is negative. For example, $\varepsilon_x = \varepsilon_z = -39.8 + 2.1i$ and $\varepsilon_y = 29.2 + 0.1i$ at $\lambda_0 = 1.55$ μm.

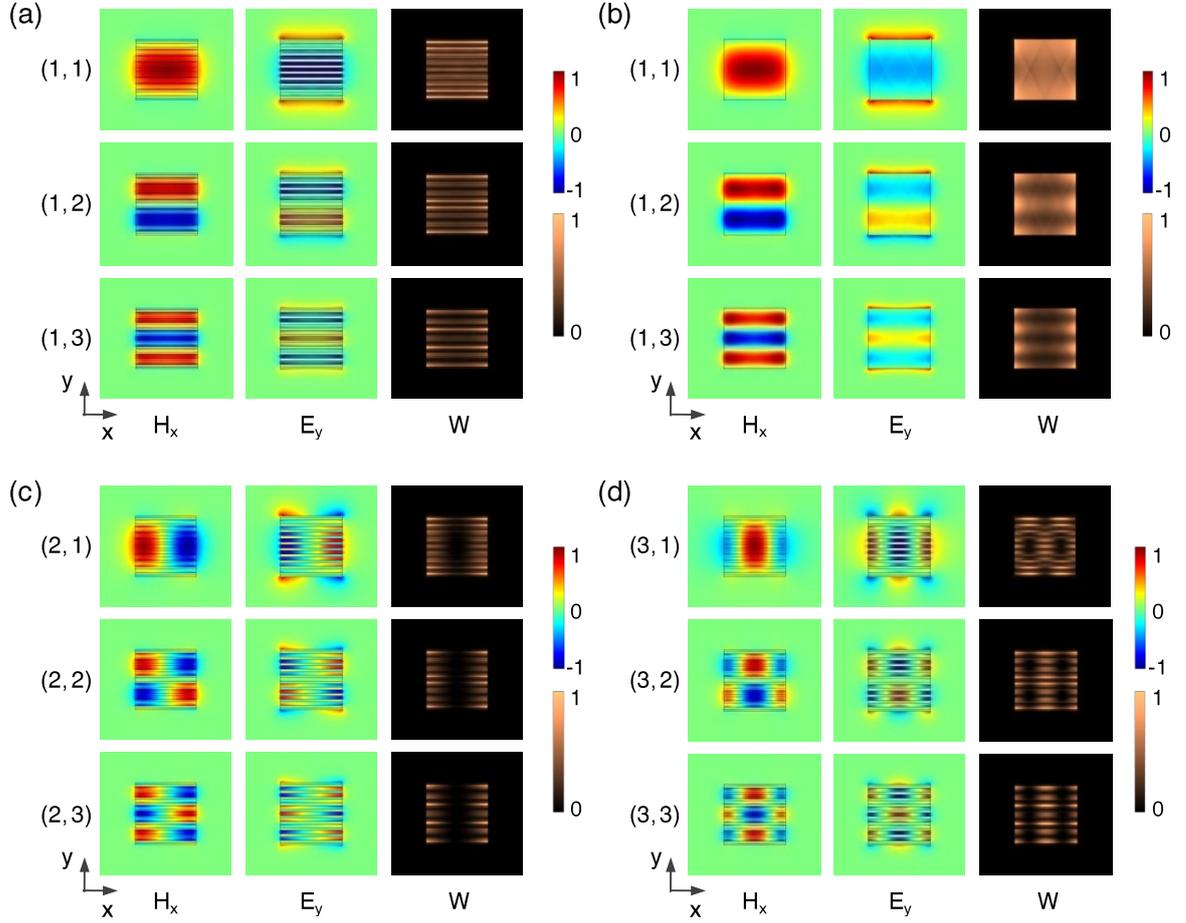

**Figure 2.** The distributions of magnetic field $H_x$, electric field $E_y$, and EM energy density $W$ for waveguide modes with different mode orders supported in the metamaterial waveguide with cross section of 100 nm by 100 nm at $\lambda_0 = 1$ μm. Waveguide modes with mode orders of (1, $m_y$), (2, $m_y$) and (3, $m_y$) in metal-dielectric multilayer waveguide structures are shown in (a), (c) and (d), respectively. The (1, $m_y$) modes are also calculated with the effective medium method, as shown in (b), which agree very well with the multilayer results in (a).

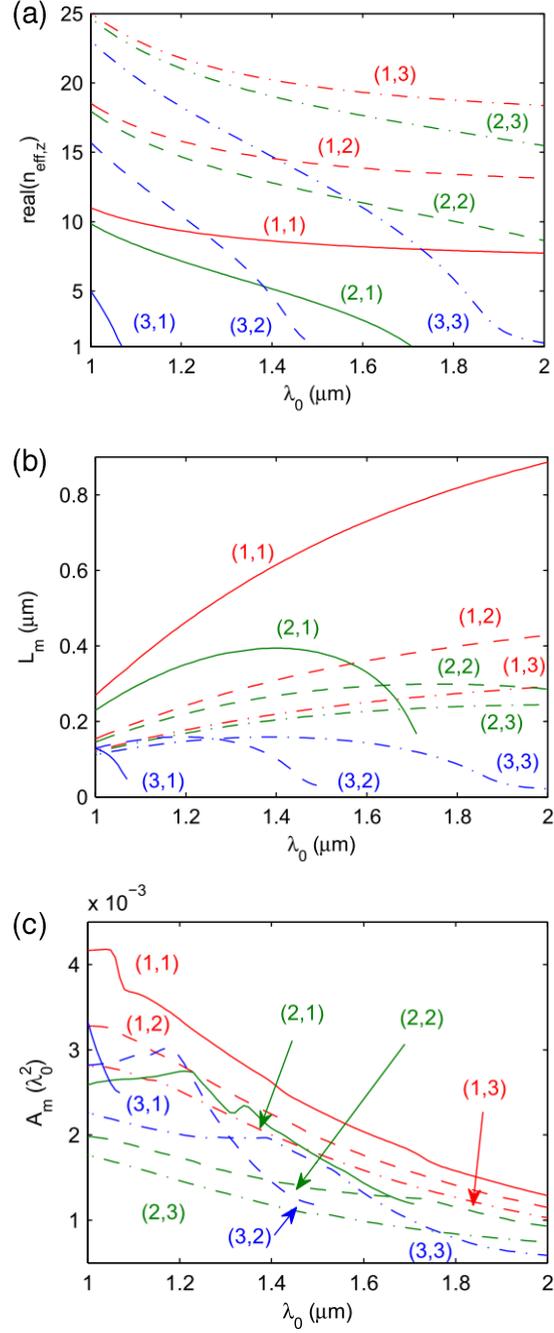

**Figure 3.** Dependences of (a) waveguide mode indices along the propagation direction $n_{\text{eff},z}$, (b) propagation lengths $L_m$ and (c) mode areas $A_m$ on wavelength $\lambda_0$ for waveguide modes with different mode orders in the waveguide with cross section of 100 nm by 100 nm. All the $(3, m_y)$ modes and the $(2, 1)$ mode have mode cutoff at certain wavelengths. The propagation lengths depend on both the absorption loss and the mode radiation loss. The mode areas are in the order of $10^{-3}$ ($\lambda_0^2$), due to the tight photon confinement of waveguide modes.

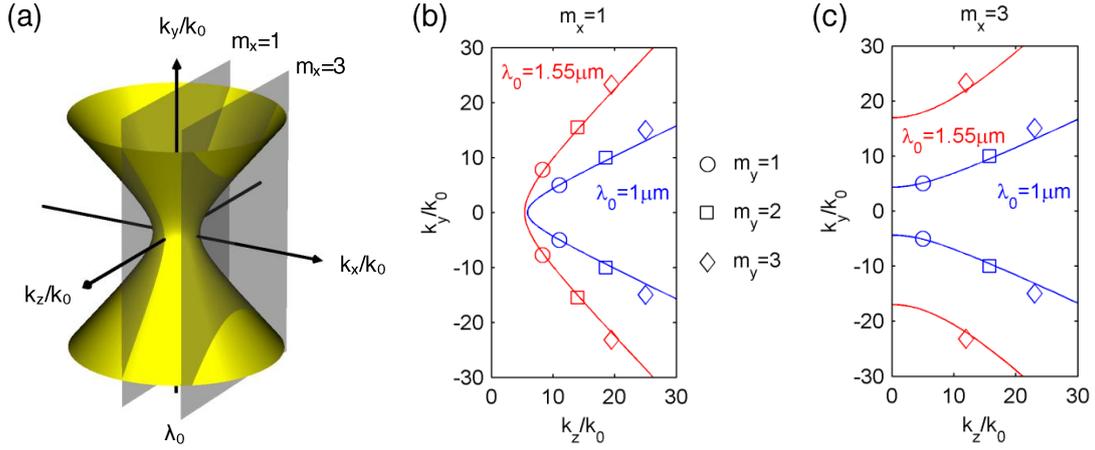

**Figure 4.** Waveguide modes plotted in *k*-space at different free space wavelengths $\lambda_0$ for the waveguide with cross section of 100 nm by 100 nm. (a) The hyperboloid IFC of the indefinite metamaterial at the wavelength of $\lambda_0$. Two cutting planes in grey color at $k_x = 0$ and $k_x/k_0 = \lambda_0/L_x$ represent mode orders of $m_x = 1$ and $m_x = 3$, respectively. The crossing curves between the two cut planes and the hyperboloid IFC are shown in (b) and (c), respectively, at different wavelengths of $\lambda_0 = 1$ μm (blue color) and $\lambda_0 = 1.55$ μm (red color). The markers in (b) and (c) represent the (1, $m_y$) and (3, $m_y$) modes calculated from multilayer metamaterial waveguide structures. It is noted that the hyperbolic curves have changed their opening directions from the *z* direction in (b) to the *y* direction in (c), which will lead to the waveguide mode cutoff for the (3, 1) and (3, 2) modes at $\lambda_0 = 1.55$ μm shown in (c). Only the $k_z > 0$ part of the IFC is shown.

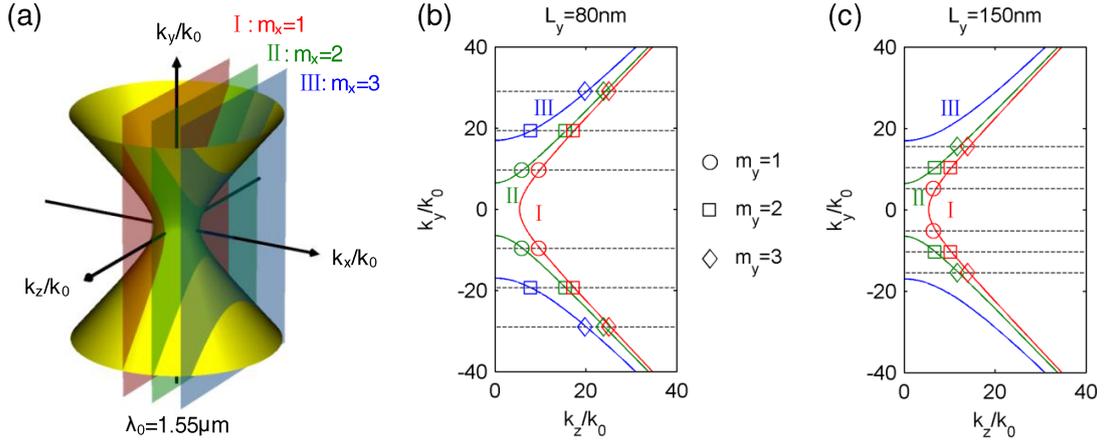

**Figure 5.** Waveguide modes plotted in *k*-space at different waveguide height $L_y$ for a fixed width $L_x$ = 100 nm. (a) The hyperboloid IFC of the indefinite metamaterial at $\lambda_0$ = 1.55 μm. Three cutting planes of I, II and III represent mode orders of $m_x = 1$, $m_x = 2$ and $m_x = 3$, respectively. The three crossing curves between the cutting planes and the hyperboloid IFC are shown in (b) and (c), for different waveguide heights of $L_y$ = 80 nm and $L_y$ = 150 nm, respectively. The markers in (b) and (c) represent the ($m_x$, $m_y$) modes calculated from multilayer metamaterial waveguide structures. The horizontal dashed lines show the location $m_y$ = 1, 2 and 3 for a certain waveguide height $L_y$. It is observed that $k_y$ will decrease as $L_y$ gets larger, which will result in the waveguide mode cutoff for high order modes, such as the (2, 1), (3, 2) and (3, 3) modes, as shown in (c). Only the $k_z > 0$ part of the IFC is shown.

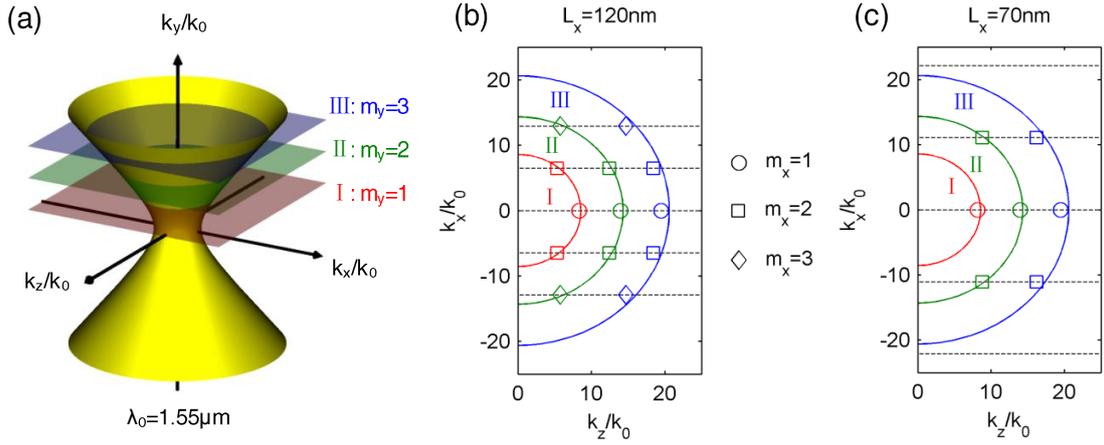

**Figure 6.** Waveguide modes plotted in *k*-space at different waveguide widths $L_x$ for a fixed height $L_y = 100$ nm. (a) The hyperboloid IFC of the indefinite metamaterial at $\lambda_0 = 1.55$ μm. Three cutting planes of I, II and III represent mode orders of $m_y = 1$, $m_y = 2$ and $m_y = 3$, respectively. The three crossing circles between the cut planes and the hyperboloid IFC are shown in (b) and (c), for different waveguide widths of $L_x = 120$ nm and $L_x = 70$ nm, respectively. The markers in (b) and (c) represent the ($m_x$, $m_y$) modes calculated from multilayer metamaterial waveguide structures. The horizontal dashed lines show the location $m_x = 1$, 2 and 3 for a certain waveguide width $L_x$. It is observed that $k_x$ will increase as $L_x$ gets smaller, which will result in the waveguide mode cutoff for high order modes, such as the (2, 1), (3, 2) and (3, 3) modes, as shown in (c). Only the $k_z > 0$ part of the IFC is shown.

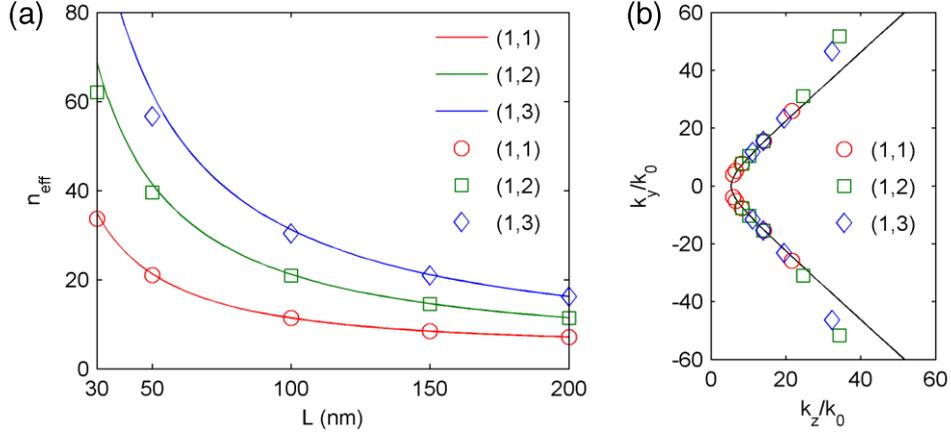

**Figure 7.** (a) Effective refractive indices $n_{\text{eff}}$ for the $(1, m_y)$ modes as functions of waveguide sizes $L$ for metamaterial waveguides with square cross sections at $\lambda_0 =$ 1.55 μm. Solid curves are calculated from the effective medium method and markers represent the simulation results from multilayer metamaterial waveguide structures. It is shown that a smaller waveguide size $L$ and a higher mode order $m_y$ will lead to a larger refractive index. (b) The wave vectors of the waveguide modes shown in (a) are plotted in $k$-space, which match the effective medium calculation (solid curve) in the low $k$ region ($k << 2\pi/a$). In the high $k$ region (when $L \leq$ 50 nm), as the wavelength is close to the period of the multilayer $a$, the wave vectors calculated from the multilayer structure deviate from the effective medium simulation at the edge of the first Brillouin zone.

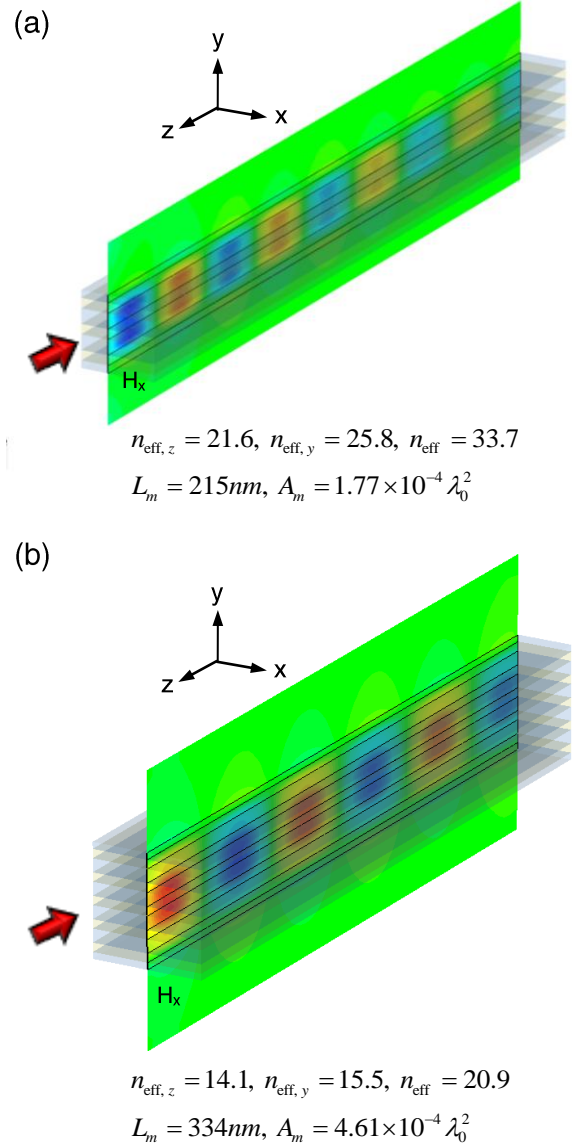

(a)

$n_{\text{eff}, z} = 21.6$, $n_{\text{eff}, y} = 25.8$, $n_{\text{eff}} = 33.7$
$L_m = 215 nm$, $A_m = 1.77 \times 10^{-4} \lambda_0^2$

(b)

$n_{\text{eff}, z} = 14.1$, $n_{\text{eff}, y} = 15.5$, $n_{\text{eff}} = 20.9$
$L_m = 334 nm$, $A_m = 4.61 \times 10^{-4} \lambda_0^2$

**Figure 8.** Mode propagation of the (1, 1) modes at $\lambda_0 = 1.55$ μm inside the metamaterial waveguides with square cross sections of (a) 30 nm by 30 nm and (b) 50 nm by 50 nm, respectively. The distributions of the magnetic field $H_x$ are plotted. The length of the waveguides in $z$ direction is 300 nm for both cases. The waveguide mode properties are listed on the bottom of each figure.